\documentclass[useAMS,usenatbib,usegraphicx]{mn2e}

\usepackage{pdflscape}

\title[]{Constraints on the circumstellar dust around KIC 8462852}
\author[M.A. Thompson]{M.A. Thompson$^{1}$\thanks{E-mail: m.a.thompson@herts.ac.uk},  P. Scicluna$^{2}$, F. Kemper$^{2}$, J.E. Geach$^{1}$, M.M. Dunham$^{3}$, \newauthor O. Morata$^{2}$, S. Ertel$^{4}$,  P.T.P. Ho$^{2,5}$,  J. Dempsey$^{5}$, I. Coulson$^{5}$,  G. Petitpas$^{3}$, \newauthor L.E. Kristensen$^{3}$\\
$^{1}$Centre for Astrophysics Research, School of Physics Astronomy \& Mathematics,
University of Hertfordshire, College Lane,\\ Hatfield, Herts, AL10 9AB, UK\\
$^{2}$Academia Sinica Institute of Astronomy and Astrophysics, P.O. Box 23-141, Taipei 10617, Taiwan\\
$^{3}$Harvard-Smithsonian Center for Astrophysics, 60 Garden Street, Cambridge, MA 02138, USA\\
$^{4}$ European Southern Observatory, Alonso de Cordova 3107, Vitacura, Casilla 19001, Santiago, Chile\\
$^{5}$East Asian Observatory, 660 N. A’ohoku Place, University Park, Hilo, Hawaii 96720, USA\\
}
\begin{document}

\date{}

\pagerange{\pageref{firstpage}--\pageref{lastpage}} \pubyear{2011}

\maketitle

\label{firstpage}

\begin{abstract}
We present millimetre (SMA) and sub-millimetre (SCUBA-2) continuum observations of the peculiar star KIC 8462852 which displayed several deep and aperiodic dips in brightness during the \emph{Kepler} mission. Our  observations are approximately confusion-limited at 850 $\mu$m and are the deepest millimetre and sub-millimetre photometry of the star that has yet been carried out. No significant emission  is detected towards KIC 8462852. We determine upper limits for dust between  a few 10$^{-6}$ M$_{\oplus}$ and 10$^{-3}$ M$_{\oplus}$ for regions identified as the most likely to host occluding dust clumps and a total overall dust budget  of $<$7.7 M$_{\oplus}$ within a radius of 200 AU.  Such low limits for the inner system make the catastrophic planetary disruption hypothesis unlikely.  Integrating over the \emph{Kepler} lightcurve we determine that at least 10$^{-9}$ M$_{\oplus}$ of dust is required to cause the observed Q16 dip. This is consistent with the currently most favoured cometary breakup hypothesis, but nevertheless implies the complete breakup of $\sim$ 30 Comet 1/P Halley type objects.  Finally, in the wide SCUBA-2 field-of-view we identify another candidate debris disc system that is potentially the largest  yet discovered.

\end{abstract}


\section{Introduction}
\label{sect:intro}

KIC 8462852 is one of the most peculiar stars discovered in the \emph{Kepler} mission, exhibiting deep and aperiodic dips in its lightcurve that are so far unexplained \citep[][hereafter B15]{boyajian2015}. 
KIC 8462852 (also known as TYC 3162-665-1 and 2MASS J20061546+4427248) is a main sequence F3 star with a possible wide M dwarf companion \citep[B15,][]{lisse2015}. Over the period of the \emph{Kepler} survey KIC 8462852 has a relatively constant brightness but during several periods the star underwent short-duration dips in brightness, including two events at a decrement of 15\% and 22\%. The lightcurves of these decreases, their durations and aperiodicity are not consistent with an origin of planetary transits.

B15 outline several possibilities for the decreases in brightness of KIC 8462852: instrumental effects; intrinsic stellar variability; catastrophic collisions of asteroids or planetary bodies; dust enshrouded planetesimals, or a family of disrupted comets. B15 examine these possibilities in detail and are able to rule out instrumental effects and several classes of intrinsically variable stars, but find that no single hypothesis fits all the known facts. The disrupted cometary family hypothesis explains the lack of periodicity in the dips in brightness (due to eccentric orbits), but requires alternative ``forward-tail'' comet geometries to explain some of the lightcurves. B15 conducted a search for similar objects in the \emph{Kepler} database and did not find any. They conclude that KIC 8462852 is almost certainly a unique phenomenon within the \emph{Kepler} database. 

\citet{wright2015} have since put forward an alternative explanation that the dips in brightness may be caused by artificial structures orbiting the star. This suggestion is motivated by the original concept of the Dyson sphere (described in \citealt{dyson1960}, although strictly first conceived by \citealt{stapledon1937}) in which a space-dwelling civilisation may tap a substantial fraction of the energy of their host star by constructing large starlight collectors in orbit around the star. Such orbiting collectors may cause transits potentially observable by \emph{Kepler} \citep{arnold2005, forgan2013}. Unsurprisingly this has provoked much speculation within the popular press on the existence of ``alien megastructures'' around the star. A  targeted SETI search for persistent microwave signals from the vicinity of KIC 8462852 has so far proved negative \citep{harp2015}.

\citet[][hereafter MHW15]{marengo2015} recently presented  warm \emph{Spitzer} photometry of the star obtained during the SpiKeS \emph{Spitzer Kepler} Survey. The \emph{Spitzer} photometry at 3.6 and 4.5 $\mu$m is a factor of a few deeper than the \emph{WISE} W1 and W2 measurements and was carried out 2 years after the dimming events observed by \emph{Kepler}. The lack of significant excess in these wavebands leads these authors to conclude that the disrupted comet hypothesis of B15 is the preferred explanation. \citet{lisse2015} find no infrared excess in a 0.8--4.2 $\mu$m spectrum of the star and reach similar conclusions. However, one of the difficulties in understanding the nature of KIC 8462852 and the existence of any circumstellar dust is the paucity of observations of the star, particularly at longer wavelengths. No sensitive photometry of the star exists at wavelengths longer than 22 $\mu$m and thus no constraint exists upon the amount of cold dust that may be present in the outer regions of the system. Emission from cold dust would peak at longer wavelengths and  be undetectable by the existing mid-infrared measurements.

In this Letter we present millimetre (mm) and sub-millimetre (sub-mm) wavelength observations of KIC 8462852 carried out with  the  Submillimeter Array and the SCUBA-2 camera on the James Clerk Maxwell Telescope. Our observations are the deepest existing mm and sub-mm photometry of  this peculiar star and allow us to place sensitive constraints on the total amount of cold dust present in the system. In Section~\ref{sect:obs} we describe our observations and present a catalogue of detected 850 $\mu$m sources. In Section \ref{sect:dustlimits} we determine upper limits to the amount of dust present in the KIC 8462852 system and discuss these limits in relation to the hypotheses presented by B15. We discuss the nature of a candidate edge-on disc discovered in the wide SCUBA-2 FOV in Sect.~\ref{sect:tyc977} which is potentially the largest debris disc yet discovered. Finally we summarise our conclusions in Section~\ref{sect:concs}. 

\section{Observations \& Results}
\label{sect:obs}

\subsection{SCUBA-2 Observations}
\label{sect:s2}

\begin{figure}
\includegraphics[width=72mm,angle=-90]{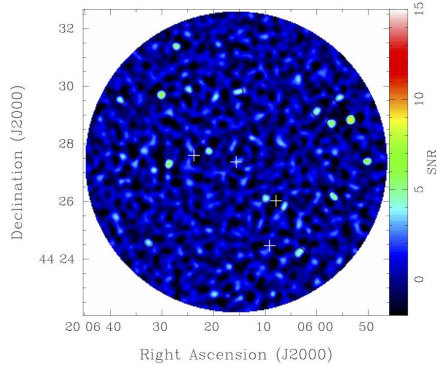}
\caption{SCUBA-2 850 $\mu$m signal to noise map of the region surrounding KIC 8462852. Sources detected in the image at greater than 4$\sigma$ confidence are indicated by circles and stars from the Tycho-2 catalogue are shown as crosses. KIC 8462852 is located at the centre of the image.}
\label{fig:daisy}
\end{figure}

We observed KIC 8462852 using the SCUBA-2 camera \citep{holland2013} mounted on the James Clerk Maxwell Telescope (JCMT) between October 26--29th 2015.  We performed our observations using the ``daisy'' mapping mode, which is optimised for mapping objects smaller than the instantaneous FOV \citep{holland2013} and results in a map of $\sim$ 13\arcmin\ diameter. Approximately 6 hours of on-source observations were obtained. Atmospheric conditions were relatively dry during our observations, with a 225 GHz zenith opacity ranging between 0.06 and 0.1. The pointing accuracy of the telescope was checked on an hourly basis and found to be accurate to within 3\arcsec. We regularly observed the calibrator CRL 2688 and estimate an absolute flux calibration of 10\% \citep{dempsey2013}.  We reduced the data with the Dynamic Iterative Map-Maker \citep{chapin2013} which is part of the \emph{Starlink} \texttt{SMURF} package \citep{jenness2013}. Given the likely compact nature of the expected emission we filtered the data at a scale of 150\arcsec\ in the map-making process to reduce the amount of low-frequency noise. A final beam matched filter was also applied to the map to enhance the sensitivity to point sources \citep{chapin2013}. The resulting 450 and 850 $\mu$m maps have sensitivities of 10.7 and 0.85 mJy respectively in a central region of 5\arcmin\ diameter. A signal to noise map of the 850 $\mu$m data is shown in Fig.~\ref{fig:daisy}.

\subsection{Submillimeter Array Observations}

One track of observations of KIC 8462852 were obtained with the Submillimeter 
Array \citep[SMA;][]{ho2004} on the 10th November 2015 in the subcompact configuration 
with eight antennas, providing projected baselines ranging from $7-72$ m.
The observations were obtained with the 345 GHz receiver at a central 
frequency of 273 GHz (1.1 mm), with a total bandwidth of 11 GHz.  The 
zenith opacity at 225 GHz was typically $\sim 0.1$ throughout the night, 
and the system temperatures ranged from 250 to 400 K depending on source 
elevation.  Regular observations of MWC 349A were interspersed with those of 
KIC 8462852 for gain calibration.  The quasar 3C273 was used for bandpass 
calibration, and Uranus was used for absolute flux calibration.  
We conservatively estimate a 20\% uncertainty in the absolute flux calibration. 
Imaging was performed with natural weighting to maximize sensitivity, giving 
a synthesized beam  of 4.4\arcsec\ $\times$ 2.8\arcsec\ at a position angle 
of 16.1$^{\rm o}$ (measured east from north).  We measure the 1$\sigma$ continuum r.m.s. 
 to be 0.73 mJy beam$^{-1}$.  

\subsection{Detected sub-mm sources in the SCUBA-2 image}
\label{sect:catalogue}

We identified sources in the SCUBA-2 850 $\mu$m image by measuring the positions of peaks with greater than 4$\sigma$ confidence in the map. A confidence level of 4$\sigma$ corresponds to a false positive rate of $\sim$5\% \citep{roseboom2013,chapin2013}, which we considered to be a reasonable level for source detection. No $\ge4\sigma$ sources were found in the 450 $\mu$m map which is fully consistent with our results at 850 $\mu$m given our 450 $\mu$m sensitivity and the typical spectral index of dust emission. The \emph{Fellwalker} algorithm  \citep{ berry2015} was used to locate the peaks. We list the detected sources, their coordinates and 850 $\mu$m fluxes in Table \ref{tbl:sources} (found in the online version of this paper). In total we detect 17 sources at 850 $\mu$m, with typical fluxes of a few mJy to 13.4 mJy. 

We find no obvious positional matches between our catalogue and the Tycho-2 or UCAC4 stellar catalogues, which implies that the majority of 850 $\mu$m sources may be background sub-mm galaxies.  The number counts of galaxies brighter than our 4$\sigma$ detection threshold are indeed consistent with extragalactic number counts \citep[10--20 galaxies brighter than this flux are predicted in an area the size of our daisy map;][]{coppin2006}. However, given the relatively low Galactic latitude of our field ($b=+6.6$) we cannot rule out a Galactic origin for at least some of the sources. In particular, we find a significant association between TYC 3162-977-1  and \emph{two} 850 $\mu$m sources  which is discussed further in Sect.~\ref{sect:tyc977}.

\section{Limits on the amount of dust associated with KIC 8462852}
\label{sect:dustlimits}

\begin{figure}
\includegraphics[width=65mm,angle=-90]{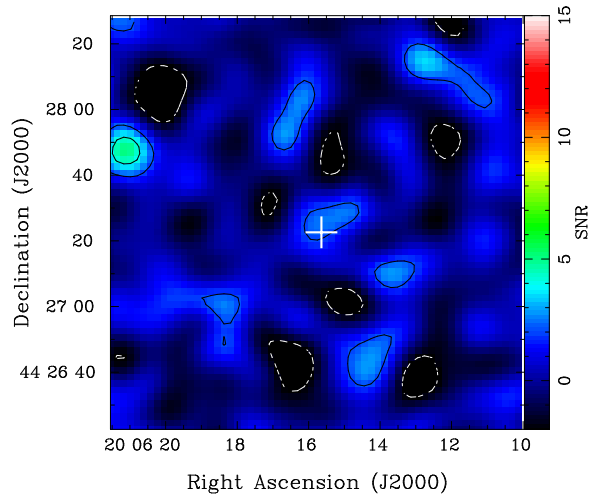}
\caption{A close-up 850 $\mu$m signal to noise image of KIC 8462852, with the star indicated by a cross symbol. Contours are plotted at $-2\sigma$ (white dashed contours), $2$ and 4$\sigma$ (black solid contours). KIC 8462852 is coincident with a 2.6$\sigma$ peak in the image, but  we do not consider this to be a robust detection.}
\label{fig:kic8462cup}
\end{figure}

\begin{figure}
\includegraphics[width=\columnwidth]{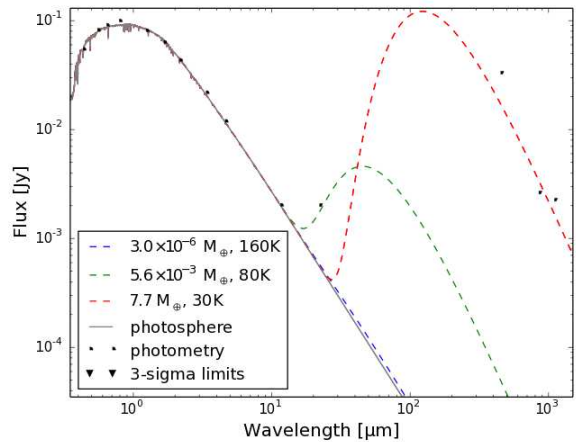}
\caption{Spectral energy distribution of  KIC 8462852 with archival photometry from B15 and MHW15 plus our SMA \& SCUBA-2 upper limits.  SED models corresponding to the scenarios discussed in Section \ref{sect:dustlimits} are also plotted.}
\label{fig:kic8462sed}
\end{figure}

We did not detect any significant emission toward KIC 8462852 in any of our observations, obtaining 3$\sigma$ upper limits at the star's position of 32.1, 2.55 and 2.19 mJy at 450, 850 and 1100 $\mu$m respectively. A 2.6$\sigma$ peak is observed at the position of KIC 8462852 in the 850 $\mu$m SCUBA-2 image (see the closeup in Fig.~\ref{fig:kic8462cup}) but we do not consider this to be a robust detection. We plot our upper limits plus the B15 and MHW15 photometry in Figure \ref{fig:kic8462sed}  together with three SED models discussed below. 

At short wavelengths the observed emission is consistent with a purely photospheric origin from the star. At longer wavelengths \emph{WISE}, SCUBA-2 and SMA  provide only upper limits to any emission from circumstellar dust. This means that it is difficult to determine an absolute mass limit of any dust that may be associated with the star as we have few constraints on the underlying geometry (and hence temperature) of the dust distribution.  However, the scenario-independent constraints proposed by B15 allow us to consider three hypothetical cases and determine the corresponding upper limits to their dust mass: \emph{i)} where dust clumps are on circular orbits and are constrained by the dip durations to lie within 2 and 8 AU from the star;  \emph{ii)} where the dust clumps lie on highly elliptical orbits and the dip durations constrain them to lie within  0.1 and 26 AU; and \emph{iii)} where the dust distribution is constrained to only be bound to the star with an outer radius of $\sim$200 AU. We consider the bound of 200 AU to be an upper limit due to the presence of  the companion to KIC 8462852.

To derive consistent upper limits on the mass of dust in the system, we compute grids of likelihoods for representative ranges of $T_{\rm dust}$ and M$_{\rm dust}$ consistent with each of our hypothetical cases, assuming that the stellar parameters are those given in B15 and that any dust emission takes the form of a modified blackbody. We assume a dust opacity $\kappa_{\nu}$ of 1.7 cm$^{2}$\,g$^{-1}$ at wavelength of 850 $\mu$m and a dust emissivity index $\beta=1$ \citep[e.g.][]{carpenter2005}. As the dust emission is constrained only by upper limits on the excess flux, the likelihood for the model fluxes at these wavelengths becomes the probability that the model would be observed to have a flux below the 3$\sigma$ limit, given the (rms-)uncertainty of the observation. By marginalising over the dust temperature, we derive upper limits to the dust mass by finding the minimum value that lies above at least 99.73\% of the integrated likelihood.




For our three hypothetical cases we find  upper limits to the dust mass for  case \emph{i)} of  3.0 $\times$ 10$^{-6}$  M$_{\oplus}$; for case  \emph{ii)} of 5.6 $\times$ 10$^{-3}$  M$_{\oplus}$; and for case \emph{iii)}  of 7.7 M$_{\oplus}$. The W3 and W4 upper limits place the tightest constraints on cases \emph{i)} and \emph{ii)}, whereas the 850 $\mu$m upper limit constrains the total amount of cold dust present in the system.

Estimating an initial mass for colliding bodies that generate a particular amount of dust in a collision is not possible, given the wide range of variables in potential collisions.
However  an upper limit for dust within 8 AU of $\sim$10$^{-6} M_{\oplus}$ makes the planetary collision hypothesis of B15 extremely unlikely. This amount of dust is comparable to  a completely pulverised object of only a tenth of the mass of 1 Ceres. Of course, as MHW15 state, if the impactors were on an elliptical orbit the obscuring dust could have since moved to a more distant point where its now much cooler spectrum is below the \emph{Spitzer} or \emph{WISE} upper limits. Our SED models preclude this hypothesis by setting stringent limits of the order of a Pluto mass of dust out to at least 26 AU, which would require around 18 years for dust in a Keplerian orbit to travel this distance. Thus any planetary mass collision generating significant amounts of dust should have been detected in the MHW15 photometry.


B15's leading hypothesis, also favoured by MHW15 and \cite{lisse2015} is that the dimming events are caused by the breakup of a comet family. We can examine this hypothesis by inferring the minimum dust mass from that required to cause the obscuration in the \emph{Kepler} light curve. We considered the quarter with the deepest absorption event (Q16) and  computed the time-dependent optical depth from the light curve, converting optical depth to dust mass by assuming the same debris-disc like dust and stellar properties used to model the SED. By integrating over this mass and correcting for the crossing time as a function of assumed velocity we arrive at a lower limit to the mass of dust required to explain the obscuration. 

For the orbits favoured by the dip duration, the resulting lower limit to the dust mass is of the order 10$^{-9}$ M$_{\oplus}$. This corresponds to a completely pulverised object with approximately 30 times the mass of  Comet 1P/Halley. 
I
It is difficult to see how this dimming event could be caused by anything other than a truly remarkable family of comets. Our simple analysis agrees remarkably well with the more complex cometary modelling of \cite{bodman2015} who estimate that the Q16 and Q17 dips require $\sim$73 comets of radius 100  km (i.e.~of the order of the radius of Comet C/1995 O1 Hale-Bopp). 

The probablity of observing a transit of a stellar radius clump of dust with a 0.1--1 AU periastron is 15--1.5\%, which implies that either there is a mechanism to provide comets or comet clusters with similar orbital inclinations (perhaps the mean motion resonances discussed by \citealt{bodman2015}), or that there is a much larger population of unseen non-transiting comets in the system. Our observations do not preclude the existence of a large Kuiper Belt or a large population of infalling comets on non-transiting orbits. However we note that there are no unusual absorption lines in the spectrum of KIC 8462852 presented by B15, which might be expected for a  population of sun-grazing comets \citep[e.g.~][]{beust1996}.

\section{TYC 3162-977-1: a potential $\sim$1000 AU edge on disc?}
\label{sect:tyc977}

\begin{figure}
\includegraphics[width=65mm,angle=-90]{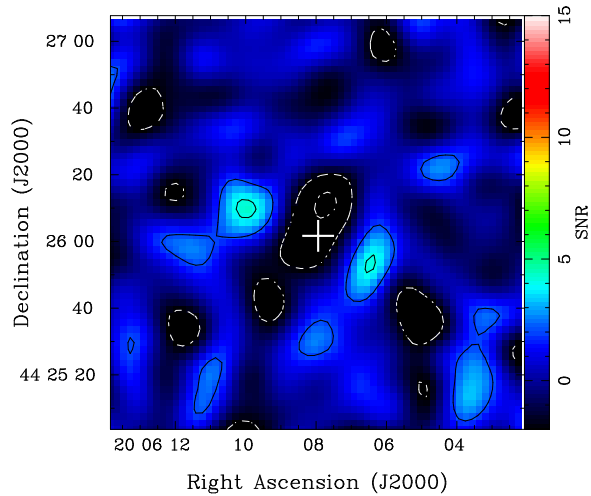}
\caption{A close-up 850 $\mu$m signal to noise image of TYC 3162-977-1, with the star indicated by a cross symbol. Contours are plotted at $-4$ and $-2\sigma$ (white dashed contours), $2$ and 4$\sigma$ (black solid contours). TYC 3162-977-1 is clearly seen in the negative bowl caused by the matched filtering process and is bracketed by two equidistant $\sim$4.3$\sigma$ 850 $\mu$m sources.}
\label{fig:edgeon}
\end{figure}

In Sect.~\ref{sect:catalogue} we did not find any stellar counterparts to single 850 $\mu$m sources. However the star TYC 3162-977-1 is equidistant from two $\sim$4 mJy 850 $\mu$m sources  (Fig.~\ref{fig:edgeon}). The star appears to reside in a significant negative dip in the map, however this is an artefact caused by the matched filter applied in the data reduction (see Sect.~\ref{sect:s2}). The overall morphology is reminiscent of the edge-on Fomalhaut system imaged with SCUBA by \citet{holland2003} and so it is possible that we are seeing a similar disc-like structure around TYC 3162-977-1.

The close correspondence of star and 850 $\mu$m sources is unlikely to result from a chance alignment. We carried out a Monte Carlo simulation of the associations between a  synthetic 1\degr\ radius Tycho-2 star catalogue corresponding to the Galactic latitude and longitude of the KIC 8462852 field and the SCUBA-2 Cosmology Legacy Survey UDS field (Geach et al.~2016, in prep). We found that there is only a 0.6\% probability of finding a pair of  850 $\mu$m sources brighter than 4 mJy within 21\arcsec\ of a star in a region the size of our daisy map. We did not restrict our simulation to colinear alignments of star and sub-mm sources, nor to pairs of sources at similar flux, and thus we consider this probability to be a strict upper limit.

TYC 3162-977-1 is classified as a K7 V dwarf by  \citet{pickles2010} and a K4 V dwarf by\citet{ammons2006}, with estimated distances of 57 and 198 pc resepectively. However, \citet{pinsonneault2012} determine a $\log\,g$ value of 2.2, which is more consistent with a subgiant or giant star than a main sequence star. For a subgiant or giant star the 850 $\mu$m emission could be explained by a reheated debris disc or perhaps a detached stellar shell, whereas for a dwarf star the best explanation is an edge-on debris disc. 

Regardless of the evolutionary state of TYC 3162-977-1, its distance, and the separation of the 850um sources imply that - if it is a debris disc (reheated or not) - then its sub-mm extent ($>$1000 AU) is by far larger than that of \emph{any} known such discs. The discs of $\beta$ Pictoris and HR8799 \citep{larwood2001, su2009} are seen to have a similar extent, but their extended emission results from small unbound dust grains blown away by radiation pressure. In contrast, at 850 $\mu$m we see larger, bound grains which are found to be in much smaller orbits in $\beta$ Pictoris and HR8799  \citep{dent2014,matthews2014}. It is noteworthy  that if $\beta$ Pictoris were moved to a similar distance to TYC 3162-977-1  its 850 $\mu$m flux \citep[][]{holland1998} would be  $\sim$4 mJy -- similar to the fluxes we measure here.

\section{Summary and Conclusions}
\label{sect:concs}

We present  450, 850 and 1100 $\mu$m continuum observations of the peculiar star KIC 8462852 carried out with the Submillimeter Array and the SCUBA-2 camera on the James Clerk Maxwell Telescope. Our observations are the deepest photometry at these wavelengths yet obtained for this star, being roughly confusion limited at 850 $\mu$m.  We detected 17 sub-mm sources in the 850 $\mu$m image but no significant  emission in the 450 \& 1100 $\mu$m images. No significant emission is detected at the position of KIC 8462852 and we obtained 3$\sigma$ upper limits at 450, 850, 1100 $\mu$m of  32.1, 2.55 and 2.19 mJy respectively. We draw the following conclusions:

\begin{enumerate} \topsep 0pt \itemsep 0 pt

\item We determine upper limits to the dust mass for three hypothetical dust geometries in the KIC 8462852 system, corresponding to the ``scenario-independent'' constraints presented by B15. There is $\le$3.0 $\times$ 10$^{-6}$ M$_{\oplus}$ of dust lying 2--8 AU from the star;  $\le$5.6 $\times$ 10$^{-3}$  M$_{\oplus}$ out to a radius of 26 AU; and an overall upper limit for dust within 200 AU of the star of 7.7 M$_{\oplus}$. These limits make the catastrophic planetary collision hypothesis of B15 extremely unlikely.

\item From integrating the opacity in the \emph{Kepler} lightcurve we obtain a lower limit to the amount of dust required to cause the deepest dip in the Q16 quarter of $\sim$10$^{-9}$ M$_{\oplus}$. This is consistent with more complex cometary models \citep{bodman2015} and requires a total amount of dust comparable to $\sim$30 completely pulverised Comet Halleys. Our photometry limits  do not rule out the existence of either a large population of infalling comets or the massive Kuiper Belt that would be their source.

\item We identify two 850 $\mu$m sources associated with the star TYC 3162-977-1 which are suggestive of an edge-on disc morphology.  The probability of this being a chance association is $\le$0.6\%.  The estimated distance to TYC 23162-977-1  implies that if this system is indeed a debris disc then it is by far the largest yet discovered with a $\sim$1000 AU radius.

\end{enumerate}

\section*{Acknowledgments}

The James Clerk Maxwell Telescope is operated by the East Asian Observatory on behalf of The National Astronomical Observatory of Japan, Academia Sinica Institute of Astronomy and Astrophysics, the Korea Astronomy and Space Science Institute, the National Astronomical Observatories of China and the Chinese Academy of Sciences (Grant No. XDB09000000), with additional funding support from the Science and Technology Facilities Council of the United Kingdom and participating universities in the United Kingdom and Canada. This work is based partly on observations obtained with the Submillimeter Array, a joint project between the Smithsonian Astrophysical Observatory and the Academia Sinica Institute of Astronomy and Astrophysics and funded by the Smithsonian Institution and the Academia Sinica. MAT acknowledges  support from the UK Science \& Technology Facility Council via grant ST/M001008/1. FK acknowledges the support the Ministry of Science and Technology of Taiwan through grant MOST104-2628-M-001-004-MY3.  JEG acknowledges the support of the Royal Society. MMD acknowledges support from the SMA through an SMA Postdoctoral Fellowship, and from NASA through grant NNX13AE54G.

\bibliography{./standardbibtex}

\begin{thebibliography}{29}
\expandafter\ifx\csname natexlab\endcsname\relax\def\natexlab#1{#1}\fi

\bibitem[{{Ammons} {et~al}\mbox{.}(2006){Ammons}, {Robinson}, {Strader},
  {Laughlin}, {Fischer}, \& {Wolf}}]{ammons2006}
{Ammons} S.~M., {Robinson} S.~E., {Strader} J., {Laughlin} G., {Fischer} D.,
  {Wolf} A., 2006, \apj, 638, 1004

\bibitem[{{Arnold}(2005)}]{arnold2005}
{Arnold} L.~F.~A., 2005, \apj, 627, 534

\bibitem[{{Berry}(2015)}]{berry2015}
{Berry} D.~S., 2015, Astronomy and Computing, 10, 22

\bibitem[{{Beust} \& {Morbidelli}(1996)}]{beust1996}
{Beust} H., {Morbidelli} A., 1996, \icarus, 120, 358

\bibitem[{{Bodman} \& {Quillen}(2015)}]{bodman2015}
{Bodman} E.~H.~L., {Quillen} A., 2015, ArXiv e-prints

\bibitem[{{Boyajian} {et~al}\mbox{.}(2015){Boyajian}, {LaCourse}, {Rappaport},
  {Fabrycky}, {Fischer}, {Gandolfi}, {Kennedy}, {Liu}, {Moor}, {Olah}, {Vida},
  {Wyatt}, {Best}, {Ciesla}, {Csak}, {Dupuy}, {Handler}, {Heng}, {Korhonen},
  {Kovacs}, {Kozakis}, {Kriskovics}, {Schmitt}, {Szabo}, {Szabo}, {Wang},
  {Goodman}, {Hoekstra}, \& {Jek}}]{boyajian2015}
{Boyajian} T.~S. {et~al.}, 2015, ArXiv e-prints

\bibitem[{{Carpenter} {et~al}\mbox{.}(2005){Carpenter}, {Wolf}, {Schreyer},
  {Launhardt}, \& {Henning}}]{carpenter2005}
{Carpenter} J.~M., {Wolf} S., {Schreyer} K., {Launhardt} R., {Henning} T.,
  2005, \aj, 129, 1049

\bibitem[{{Chapin} {et~al}\mbox{.}(2013){Chapin}, {Berry}, {Gibb}, {Jenness},
  {Scott}, {Tilanus}, {Economou}, \& {Holland}}]{chapin2013}
{Chapin} E.~L., {Berry} D.~S., {Gibb} A.~G., {Jenness} T., {Scott} D.,
  {Tilanus} R.~P.~J., {Economou} F., {Holland} W.~S., 2013, \mnras, 430, 2545

\bibitem[{{Coppin} {et~al}\mbox{.}(2006){Coppin}, {Chapin}, {Mortier}, {Scott},
  {Borys}, {Dunlop}, {Halpern}, {Hughes}, {Pope}, {Scott}, {Serjeant}, {Wagg},
  {Alexander}, {Almaini}, {Aretxaga}, {Babbedge}, {Best}, {Blain}, {Chapman},
  {Clements}, {Crawford}, {Dunne}, {Eales}, {Edge}, {Farrah}, {Gazta{\~n}aga},
  {Gear}, {Granato}, {Greve}, {Fox}, {Ivison}, {Jarvis}, {Jenness}, {Lacey},
  {Lepage}, {Mann}, {Marsden}, {Martinez-Sansigre}, {Oliver}, {Page},
  {Peacock}, {Pearson}, {Percival}, {Priddey}, {Rawlings}, {Rowan-Robinson},
  {Savage}, {Seigar}, {Sekiguchi}, {Silva}, {Simpson}, {Smail}, {Stevens},
  {Takagi}, {Vaccari}, {van Kampen}, \& {Willott}}]{coppin2006}
{Coppin} K. {et~al.}, 2006, \mnras, 372, 1621

\bibitem[{{Dempsey} {et~al}\mbox{.}(2013){Dempsey}, {Friberg}, {Jenness},
  {Tilanus}, {Thomas}, {Holland}, {Bintley}, {Berry}, {Chapin}, {Chrysostomou},
  {Davis}, {Gibb}, {Parsons}, \& {Robson}}]{dempsey2013}
{Dempsey} J.~T. {et~al.}, 2013, \mnras, 430, 2534

\bibitem[{{Dent} {et~al}\mbox{.}(2014){Dent}, {Wyatt}, {Roberge}, {Augereau},
  {Casassus}, {Corder}, {Greaves}, {de Gregorio-Monsalvo}, {Hales}, {Jackson},
  {Hughes}, {Lagrange}, {Matthews}, \& {Wilner}}]{dent2014}
{Dent} W.~R.~F. {et~al.}, 2014, Science, 343, 1490

\bibitem[{{Dyson}(1960)}]{dyson1960}
{Dyson} F.~J., 1960, Science, 131, 1667

\bibitem[{{Forgan}(2013)}]{forgan2013}
{Forgan} D.~H., 2013, Journal of the British Interplanetary Society, 66, 144

\bibitem[{{Harp} {et~al}\mbox{.}(2015){Harp}, {Richards}, {Shostak}, {Tarter},
  {Vakoch}, \& {Munson}}]{harp2015}
{Harp} G.~R., {Richards} J., {Shostak} S., {Tarter} J.~C., {Vakoch} D.~A.,
  {Munson} C., 2015, ArXiv e-prints

\bibitem[{{Ho}, {Moran} \& {Lo}(2004){Ho}, {Moran}, \& {Lo}}]{ho2004}
{Ho} P.~T.~P., {Moran} J.~M., {Lo} K.~Y., 2004, \apjl, 616, L1

\bibitem[{{Holland} {et~al}\mbox{.}(2013){Holland}, {Bintley}, {Chapin},
  {Chrysostomou}, {Davis}, {Dempsey}, {Duncan}, {Fich}, {Friberg}, {Halpern},
  {Irwin}, {Jenness}, {Kelly}, {MacIntosh}, {Robson}, {Scott}, {Ade},
  {Atad-Ettedgui}, {Berry}, {Craig}, {Gao}, {Gibb}, {Hilton}, {Hollister},
  {Kycia}, {Lunney}, {McGregor}, {Montgomery}, {Parkes}, {Tilanus}, {Ullom},
  {Walther}, {Walton}, {Woodcraft}, {Amiri}, {Atkinson}, {Burger}, {Chuter},
  {Coulson}, {Doriese}, {Dunare}, {Economou}, {Niemack}, {Parsons},
  {Reintsema}, {Sibthorpe}, {Smail}, {Sudiwala}, \& {Thomas}}]{holland2013}
{Holland} W.~S. {et~al.}, 2013, \mnras, 430, 2513

\bibitem[{{Holland} {et~al}\mbox{.}(2003){Holland}, {Greaves}, {Dent}, {Wyatt},
  {Zuckerman}, {Webb}, {McCarthy}, {Coulson}, {Robson}, \&
  {Gear}}]{holland2003}
---, 2003, \apj, 582, 1141

\bibitem[{{Holland} {et~al}\mbox{.}(1998){Holland}, {Greaves}, {Zuckerman},
  {Webb}, {McCarthy}, {Coulson}, {Walther}, {Dent}, {Gear}, \&
  {Robson}}]{holland1998}
---, 1998, \nat, 392, 788

\bibitem[{{Jenness} {et~al}\mbox{.}(2013){Jenness}, {Chapin}, {Berry}, {Gibb},
  {Tilanus}, {Balfour}, {Tilanus}, \& {Currie}}]{jenness2013}
{Jenness} T., {Chapin} E.~L., {Berry} D.~S., {Gibb} A.~G., {Tilanus} R.~P.~J.,
  {Balfour} J., {Tilanus} V., {Currie} M.~J., 2013, {SMURF: SubMillimeter User
  Reduction Facility}. Astrophysics Source Code Library

\bibitem[{{Larwood} \& {Kalas}(2001)}]{larwood2001}
{Larwood} J.~D., {Kalas} P.~G., 2001, \mnras, 323, 402

\bibitem[{{Lisse}, {Sitko} \& {Marengo}(2015){Lisse}, {Sitko}, \&
  {Marengo}}]{lisse2015}
{Lisse} C.~M., {Sitko} M.~L., {Marengo} M., 2015, ArXiv e-prints

\bibitem[{{Marengo}, {Hulsebus} \& {Willis}(2015){Marengo}, {Hulsebus}, \&
  {Willis}}]{marengo2015}
{Marengo} M., {Hulsebus} A., {Willis} S., 2015, \apjl, 814, L15

\bibitem[{{Matthews} {et~al}\mbox{.}(2014){Matthews}, {Kennedy}, {Sibthorpe},
  {Booth}, {Wyatt}, {Broekhoven-Fiene}, {Macintosh}, \&
  {Marois}}]{matthews2014}
{Matthews} B., {Kennedy} G., {Sibthorpe} B., {Booth} M., {Wyatt} M.,
  {Broekhoven-Fiene} H., {Macintosh} B., {Marois} C., 2014, \apj, 780, 97

\bibitem[{{Pickles} \& {Depagne}(2010)}]{pickles2010}
{Pickles} A., {Depagne} {\'E}., 2010, \pasp, 122, 1437

\bibitem[{{Pinsonneault} {et~al}\mbox{.}(2012){Pinsonneault}, {An},
  {Molenda-{\.Z}akowicz}, {Chaplin}, {Metcalfe}, \&
  {Bruntt}}]{pinsonneault2012}
{Pinsonneault} M.~H., {An} D., {Molenda-{\.Z}akowicz} J., {Chaplin} W.~J.,
  {Metcalfe} T.~S., {Bruntt} H., 2012, \apjs, 199, 30

\bibitem[{{Roseboom} {et~al}\mbox{.}(2013){Roseboom}, {Dunlop}, {Cirasuolo},
  {Geach}, {Smail}, {Halpern}, {van der Werf}, {Almaini}, {Arumugam}, {Asboth},
  {Auld}, {Blain}, {Bremer}, {Bock}, {Bowler}, {Buitrago}, {Chapin}, {Chapman},
  {Chrysostomou}, {Clarke}, {Conley}, {Coppin}, {Danielson}, {Farrah}, {Glenn},
  {Hatziminaoglou}, {Ibar}, {Ivison}, {Jenness}, {van Kampen}, {Karim},
  {Mackenzie}, {Marsden}, {Meijerink}, {Micha{\l}owski}, {Oliver}, {Page},
  {Pearson}, {Scott}, {Simpson}, {Smith}, {Spaans}, {Swinbank}, {Symeonidis},
  {Targett}, {Valiante}, {Viero}, {Wang}, {Willott}, \&
  {Zemcov}}]{roseboom2013}
{Roseboom} I.~G. {et~al.}, 2013, \mnras, 436, 430

\bibitem[{{Stapledon}(1937)}]{stapledon1937}
{Stapledon} O., 1937, {Star Maker}. Methuen

\bibitem[{{Su} {et~al}\mbox{.}(2009){Su}, {Rieke}, {Stapelfeldt}, {Malhotra},
  {Bryden}, {Smith}, {Misselt}, {Moro-Martin}, \& {Williams}}]{su2009}
{Su} K.~Y.~L. {et~al.}, 2009, \apj, 705, 314

\bibitem[{{Wright} {et~al}\mbox{.}(2015){Wright}, {Cartier}, {Zhao},
  {Jontof-Hutter}, \& {Ford}}]{wright2015}
{Wright} J.~T., {Cartier} K.~M.~S., {Zhao} M., {Jontof-Hutter} D., {Ford}
  E.~B., 2015, ArXiv e-prints

\end{thebibliography}
\bibliographystyle{mn2e}

\bsp

\appendix\section{Online Material}

\begin{table}
\caption{Positions, fluxes and signal to noise ratios of sources detected in the 850 $\mu$m SCUBA-2 map.}
\label{tbl:sources}
\begin{tabular}{|l|l|l|r|r|}
\hline
  \multicolumn{1}{|c|}{Source} &
  \multicolumn{1}{c|}{R.A.} &
  \multicolumn{1}{c|}{Dec.} &
  \multicolumn{1}{c|}{S$_{850}$} &
  \multicolumn{1}{c|}{SNR} \\
  \multicolumn{1}{|c|}{ID} &
  \multicolumn{1}{c|}{(J2000)} &
  \multicolumn{1}{c|}{(J2000)} &
  \multicolumn{1}{c|}{(mJy)} &
  \multicolumn{1}{c|}{} \\
\hline
  SMM 1 & 20:05:53.3 & +44:28:53 & 13.4 & 9.3\\
  SMM 2 & 20:05:50.0 & +44:27:26 & 11.3 & 6.8\\
  SMM 3 & 20:06:30.0 & +44:29:45 & 7.9 & 6.6\\
  SMM 4 & 20:05:56.9 & +44:28:45 & 8.1 & 6.0\\
  SMM 5 & 20:06:28.7 & +44:27:19 & 6.5 & 5.9\\
  SMM 6 & 20:06:27.0 & +44:31:25 & 7.7 & 5.6\\
  SMM 7 & 20:05:56.5 & +44:26:13 & 8.1 & 5.4\\
  SMM 8 & 20:06:20.8 & +44:27:47 & 4.6 & 5.0\\
  SMM 9 & 20:06:38.0 & +44:29:35 & 6.7 & 4.6\\
  SMM 10 & 20:06:03.3 & +44:24:15 & 6.7 & 4.5\\
  SMM 11 & 20:05:59.7 & +44:29:11 & 5.7 & 4.5\\
  SMM 12 & 20:05:55.4 & +44:29:39 & 6.1 & 4.5\\
  SMM 13 & 20:06:15.2 & +44:29:51 & 5.4 & 4.4\\
  SMM 14 & 20:06:09.6 & +44:26:09 & 4.6 & 4.3\\
  SMM 15 & 20:06:06.1 & +44:25:53 & 4.9 & 4.1\\
  SMM 16 & 20:06:32.4 & +44:24:37 & 5.4 & 4.0\\
  SMM 17 & 20:06:24.7 & +44:29:57 & 4.7 & 4.0\\
\hline\end{tabular}
\end{table}

\label{lastpage}

\end{document}